\documentclass[doublecol]{epl2} 

\title{ Revisiting the symmetric reactions for synthesis of super heavy nuclei of $Z\geq $120}

\author{R. K. ~Choudhury \and Y. K. Gupta}

\institute{                    
  Nuclear Physics Division, Bhabha Atomic Research Centre,Trombay, Mumbai, 400085, India\\
}
\pacs{25.70.Jj}{Fusion and fusion-fission reactions}
\pacs{25.70.Gh}{Compound nucleus}

\abstract{
Extensive efforts have been made experimentally to reach nuclei in the super heavy mass region of Z = 110 and
above with suitable choices of projectile and target nuclei. The cross sections for production of these nuclei are
seen to be in the range of a few picobarn or less, and pose
great experimental challenges. Theoretically, there have
been extensive calculations for highly asymmetric (hot-fusion) and
moderately asymmetric (cold-fusion) collisions and only a few
theoretical studies are available for near symmetric collisions
to estimate the cross sections for production of super-heavy
nuclei. In the present article, we revisit the symmetric heavy ion reactions
with suitable combinations of projectile and target nuclei in the
rare-earth region, that will lead to compound systems with very low excitation energy and with better neutron-to-proton ratio
for higher stability.}

\begin{document}

\maketitle

\section{Introduction}
An island of super heavy nuclei, with half lives ranging from a
few seconds to a few thousands of years has been predicted by
calculations based on macroscopic-microscopic theories
\cite{Mayer1966, Mosel-Greiner1968-1969, NILSSON1968-1969, NILSSON1969, JR_Nix1972}. The
large stability arises due to strong shell effects in the range of
proton numbers ($Z$ = 114 - 126) and neutron numbers ($N$  = 170 -
188), which in turn gives rise to large fission barriers (5 - 8
MeV) in this mass region. There have been extensive efforts
experimentally to synthesize super-heavy elements (SHE) through
heavy ion reactions with suitable choice of projectile and target
nuclei. However, the compound nuclei are formed in the excitation energy of few tens of MeV, and due to washing out of
shell effects with increasing excitation energy, the production
cross sections are usually quite low (in the range of picobarn
or less) for compound nuclei with $Z$ = 110 and above.
Nevertheless, nuclei with $Z$ up to 118 have been synthesized in
laboratory by various experiments so far \cite{Oganessian_nature,
Oganessian1999, Oganessian2004_69, Oganessian2004_70,
Oganessian2005_72, Oganessian2006_74, Oganessian2010, berkely2009,  hofman1998, hofman2000, hofman2004}. The
two main routes followed are: `hot fusion' with actinide target
nuclei and highly asymmetric reaction channels  \cite{Oganessian_nature, Oganessian1999, Oganessian2004_69,
Oganessian2004_70, Oganessian2005_72, Oganessian2006_74,
Oganessian2010, berkely2009}, and `cold fusion' with Pb, Bi target nuclei with
moderately asymmetric reaction channels \cite{hofman1998, hofman2000, hofman2004}. In all these experiments, the
compound nucleus (CN) is formed with relatively less neutron numbers as
compared to that needed for the extra stability due to the shell
effects.

Theoretically, there have been many attempts to understand the
reaction mechanism leading to the formation of the super heavy
nuclei \cite{swiatecki2004, swiatecki2005,Wilczynska2007,Zagrebaev2008, cap2011,Liang2012,smolancznk2001}. Based upon the
various theoretical formalisms, different reaction routes for the synthesis
of super heavy nuclei have been proposed \cite{Denisov2001, Zagrebaev2010, Iwamoto1996, Norenberg1994-1,Norenberg1994-2,Aritomo1996}. The main considerations in selecting a reaction channel for producing super-heavy nuclei are the following:
\begin{table*}
\caption{\label{tab:table1} Relevant data for the new reaction routes using the rare-earth nuclei.}
\begin{center}
\begin{tabular}{cccccc}
\hline
Reaction                            & $Z_{P}Z_{T}$  & g.s. deformations     & $Q$- Value &$V_\mathrm{Coul}$  & $S_{n}$\\
 ($Z_\mathrm{CN}$\ $A_\mathrm{CN}$) &               & (Projectile, Target)  & (MeV)      &  (MeV)            & (MeV)  \\
\hline
 $^{154}$Sm + $^{150}$Nd     &3720    &  (0.27, 0.24)      & -377.5     & 373.9       & 7.1\\\
       (122, 304)                  &                   &            &             &&\\\
$^{154}$Sm + $^{154}$Sm      &3844     & (0.27, 0.27)       &-394.9      &385.5        &7.1\\
 (124, 308)                        &                   &            &             & &\\
 $^{160}$Gd + $^{154}$Sm      &3968     &  (0.28, 0.27)      & -412.2     &  396.2      &   7.3\\\
      (126, 314)                    &                   &            &             & &\\
\hline
\hline
\end{tabular}
\end{center}
\end{table*}

\begin{enumerate}
\item{Large fusion cross section.}
\item{Low excitation energy  of CN for optimum survival probability.}
\item{Proper neutron-to-proton ratio ($n/p$) of CN for better stability.}
\item{High beam intensity and target concentration for good yield.}
\end{enumerate}

One of the main reasons for poor success of the experiments is
that the reaction $|Q|$-value is much lower than (for
`hot-fusion'), or similar to (for `cold-fusion') the Coulomb
barrier ($V_\mathrm{Coul}$) of the fusing target and projectile
nuclei. Hence, at beam energies just above Coulomb barrier, the CN
is formed with high excitation energy which is already larger than
the neutron emission threshold. For example, for
$^{48}$Ca + $^{249}$Cf [$(Z, A)_\mathrm{CN}$ = (118, 297)],  $Q$=
-174.48 MeV, $V_\mathrm{Coul}$ = 205.4 MeV and for $^{76}$Ge +
$^{208}$Pb [$(Z, A)_\mathrm{CN}$ = (114, 284)],  $Q$= -260.25 MeV,
$V_\mathrm{Coul}$ = 272.1 MeV. Similar is the case for other
target-projectile combinations of hot- and cold-fusion reactions
being used for the SHE synthesis.
\begin{figure*}[t]
\begin{center}
\centering\includegraphics [trim= 0.1mm 0.1mm 0.2mm  0.5mm, angle=360, clip, height=0.25\textheight]
{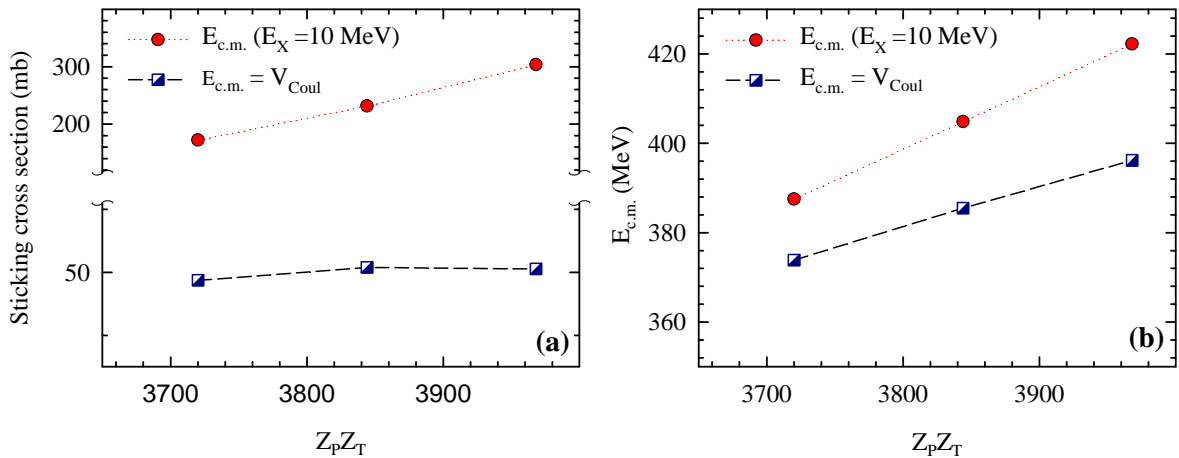}
\caption { (Color online) (a) Sticking cross section at two
different $E_{c.m.}$ values as a function of $Z_{P}Z_{T}$ for the
reactions discussed in the present work. Solid circles are for
$E_{c.m.}$ values for which $E_{X}$ =10 MeV and squares are for
$E_{c.m.}$ = $V_\mathrm{Coul}$, corresponding  center-of-mass
energies as a function of $Z_{P}Z_{T}$ in (b).  The lines in (a)
and (b) are shown to guide the eye.} \label{stick}
\end{center}
\end{figure*}

As mentioned above, the cross-sections for SHE production have
been found to be in the range of only a few picobarn or less in
the experiments carried out so far. Recent reviews \cite{Zagrebaev2008,Greiner2012, hofman2001, Ackermann2005} have emphasized
on radioactive-ion-beam routes for producing
$Z_\mathrm{CN}\geq 120$. In order to have better survival
probability, radioactive neutron rich beams ( $^{96}$Sr,
$^{132}$Sn) are being suggested to reach a more suitable neutron/
proton combination. However, these reactions will have severe
limitation on beam intensity.

\section{Symmetric heavy-ion collisions using rare-earth nuclei}
There have been some attempts using nearly symmetric collisions
such as $^{136}$Xe + $^{136}$Xe to synthesize Hs nuclei for which upper limit of the production cross section was obtained to be 4 picobarn  \cite{Oganessian2009}. The symmetric collisions using deformed projectile and target nuclei have also been suggested earlier 
\cite{Iwamoto1996, Norenberg1994-1,Norenberg1994-2} to 
synthesize super heavy nuclei. For  $^{149}$La  + $^{149}$La collision to produce $Z$=114 nuclei, the upper limit of cross section was estimated from theoretical consideration to be around 10 picobarn \cite{Aritomo1996}. However, there is no experimental data available for this system. 
In the following, we revisit the near symmetric collisions involving rare-earth nuclei
that might prove useful for synthesis of cold super-heavy nuclei.

Table \ref{tab:table1} shows some relevant data such as the
$Z_{P}Z_{T}$ value, ground state (g.s.) deformations of projectile
and target nuclei (from Ref. \cite{Moller1995}), the fusion
$Q$-value, $V_\mathrm{Coul}$ and the neutron separation energy
($S_{n}$) for certain reaction routes using rare-earth nuclei
fusion channels. The $Q$ and $S_{n}$ values are calculated using
the predicted masses by M\"{o}ller and Nix \cite{Moller1995}. The
$V_\mathrm{Coul}$ values are taken from the NRV code \cite{nrv}
which are consistent with the parameterizations of mean value of
the barrier distribution given in Ref. \cite{cap2011}. In addition
to the reactions shown in the Table \ref{tab:table1}, many more
fusion reaction channels are feasible using other different
rare-earth target/ projectile combinations. The advantages that
these reactions offer are:
\begin{figure}
\begin{center}
\centering\includegraphics [trim= 0.1mm 0.1mm 0.1mm  0.1mm, angle=360, clip, height=0.22\textheight]
{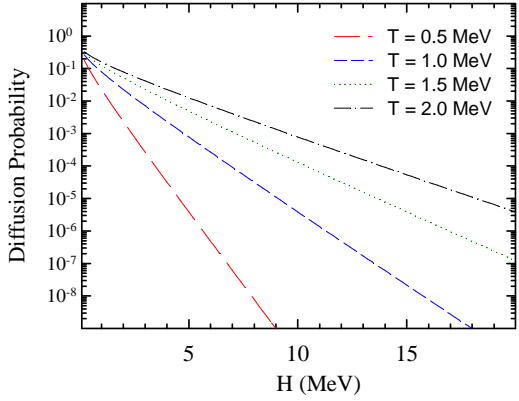}
\caption { (Color online) Diffusion probability as a function of
barrier height, $H$ opposing fusion along the asymmetric fission
valley, as seen from the injection point. Different lines
correspond to the different   temperature values (see text). } \label{diff_prob}
\end{center}
\end{figure}
\begin{figure}
\begin{center}
\centering\includegraphics [trim= 0.0mm 0.0mm 0.0mm  0.0mm, angle=360, clip, height=0.22\textheight]{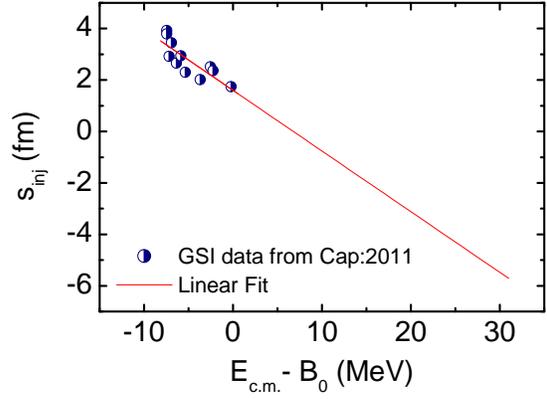}
\caption { The injection parameter ($s_{inj}$) as a function of
($E_{c.m.} - B_{0}$) taken from Ref. \cite{cap2011}). The solid
line is the least square linear fit, $s_\mathrm{inj} = 1.5985 -
0.23587 (E_{c.m.} - B_{0})$ fm/MeV.  } \label{sinj_Ecm_B0}
\end{center}
\end{figure}
\begin{enumerate}
\item{  $V_\mathrm{Coul}< |Q|$ value.}
\item{Large g.s. deformations of both target and projectile nuclei that might  enhance  near barrier fusion cross section by channel coupling and lowering of fusion barrier, $B_{fus}$.}
\item{Good $n/p$ ratio of CN.}
\item{Stable beams for large beam intensity.}
\item{Large elemental abundances of rare-earth elements.}
\item{Large center-of-mass velocity for better collection of CN residues in forward direction.}
\item{Low neutron background at optimum low bombarding energy.}
\end{enumerate}

For example, in case of  $^{160}$Gd + $^{154}$Sm   reaction, the
CN is (126, 314) where $V_\mathrm{Coul}$ is 16 MeV lower than the
energy required ($|Q|$-value) for initiating the reaction.  With
optimum above barrier bombarding energy, the CN can be produced with relatively
low excitation energy.


\section{\label{sec:level3}    Theoretical  estimates}
We will now describe some method to calculate the fusion/survival
probability of the above rare-earth reaction channels. One expects
that due to large $Z_{P}Z_{T}$ product, fusion will be
largely hindered.  However,  for deformed nuclei  there is no clear cut understanding
of the fusion hindrance (except the extra push
effects suggested by W. Swiatecki \cite{swiatecki1981,
swiatecki1982}). There are calculations reported in
literature, where only target deformation is considered
\cite{Liang2012}. We  discuss below the basic
method to have approximate estimates for the formation cross
section of the super-heavy nuclei using rare-earth nuclear
collisions.

In case of heavy colliding systems typically used for super-heavy
mass-region, overcoming the Coulomb barrier is not enough to form
the super-heavy compound nucleus. There are two avenues for
estimating the compound nuclear formation cross section for heavy
colliding nuclei similar to the ones discussed in the present
article. These are: (i) extra-extra push model
\cite{swiatecki1981, swiatecki1982} and (ii) Fusion by Diffusion
model (FBD) \cite{swiatecki2005}. According to the
extra-extra-push model, an extra energy (`extra-extra-push') with
respect to the Coulomb barrier is needed to land inside the
unconditional saddle point which guards the colliding system
against re-separation before forming the compound nucleus. The
`extra-extra-push' energy increases rapidly with effective
fissility, given by \cite{swiatecki1982};
\begin{equation}
\chi_{eff}=\frac{(Z^2/A)_{eff}}{(Z^2/A)_{crit}}
\label{eq:one}
\end{equation}

For the present reactions where deformed projectile and
target nuclei are considered, due to broad Coulomb barrier
distribution, a large amount of `extra-extra-push' energy could be
available ($\sim$100 MeV) at very low probability for certain orientations of the
colliding deformed nuclei. However, the
entrance channel barrier distributions for these kind of heavy
deformed nuclei with inclusion of dynamical effects are not easily
calculable.

On the other hand, the FBD model has been successfully
employed in reproducing the measured excitation function of the
super-heavy element synthesis \cite{swiatecki2005}. A set of twelve fusion reactions has been analyzed with the original
version of the FBD model by Swiatecki et al. \cite{swiatecki2005}. With the improved version of the FBD model, the experimental excitation functions of a complete set of 27 cold fusion reactions have been reproduced by Cap et al.~\cite{cap2011}. 
In the FBD
model, the evaporation residue cross section $\sigma_{ER}$ for
production of a given final nucleus in its ground state is
factorized as the product of the partial sticking cross-section
$\sigma_{stick}(\ell)$, the diffusion probability
$P_\mathrm{Diffus}(\ell)$, and the survival probability
$P_{surv}(\ell)$ \cite{cap2011}:
\begin{equation}
\sigma_{ER}= \sum_{\ell=0}^{\infty} \sigma_{stick}(\ell) P_\mathrm{Diffus}(\ell) P_{surv}(\ell)
\label{eq:two}
\end{equation}

\begin{equation}
~~~~~= \frac{\pi \hbar ^{2}}{2\mu E_\mathrm{c.m.}}\sum_{\ell=0}^{\ell_\mathrm{max}} (2\ell + 1) P_\mathrm{Diffus}(\ell) P_{surv}(\ell)
\label{eq:three}
\end{equation}
By replacing the summation in above equation by an integral, one obtains the sticking cross section as:
\begin{equation}
\sigma_{stick}=\frac{\pi \hbar ^{2}}{2\mu E_\mathrm{c.m.}} (\ell_\mathrm{max}+1)^{2}
\label{eq:four}
\end{equation}
where $\ell_\mathrm{max}$ is determined by the ``diffused barrier
formula'' based on assumption of Gaussian distribution of the
barriers around a mean value $B_{0}$ (see Ref. \cite{cap2011} for
details). The sticking cross sections determined for the present
reactions are shown in the Fig. \ref{stick} (a) as a function of
$Z_{P}Z_{T}$ at two center-of-mass energies: (i) $E_{c.m.}$ =
$V_\mathrm{Coul}$ and (ii) $E_{c.m.}$ values for which initial
excitation energy of the CN, $E_{X}$ =10 MeV.  Corresponding
center-of-mass energies as a function of $Z_{P}Z_{T}$ are shown in
the Fig. \ref{stick} (b).  The difference between the $E_{c.m.}$ values for
$E_{c.m.}$ = $V_\mathrm{Coul}$ and $E_{X}$ =10 MeV increases with
$Z_{P}Z_{T}$ which is reflected in the behavior of sticking cross
section as a function of $Z_{P}Z_{T}$. The lines in Fig.
\ref{stick} (a) and (b) are shown to guide the eye.

\begin{figure}
\begin{center}
\centering\includegraphics [trim= 0.0mm 0.0mm 0.0mm  -1.0mm, angle=360, clip, height=0.25\textheight]{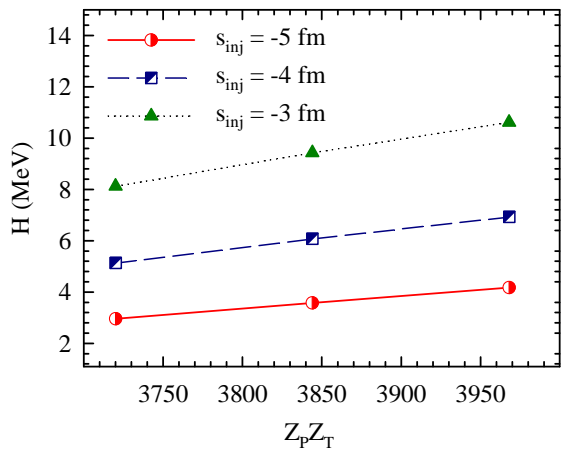}
\caption {(Color online) The barrier height ($H$) as a function of
$Z_{P}Z_{T}$ of the present reactions at various values of
injection parameter ($s_{inj}$). } \label{H_ZPZT}
\end{center}
\end{figure}

In the FBD model, the probability ($P_\mathrm{Diffus}$) that the
system injected at a point   outside the saddle point achieves
fusion is calculated using the diffusion process over a parabolic
barrier \cite{swiatecki2005}. If $L$ stands for the total length
of di-nuclear shape, the parameter $s$ is defined as $s = L -
2(R_{1} + R_{2})$. In the entrance channel of two approaching
nuclei, $s=0$ would correspond to contact of half density
contours. The  diffusion probability $P_\mathrm{Diffus}$ is then
given by \cite{cap2011, swiatecki2005}:
\begin{equation}
P_\mathrm{Diffus} = \frac{1}{2} \left (1 - erf \sqrt{H/T}\right)
\label{eq:five}
\end{equation}
where $H$ is barrier height opposing fusion along the asymmetric
fission valley, as seen from the injection point ($s_{inj}$) and
$T$ is the temperature of the fusing system. The diffusion
probability as a function of barrier height, $H$ at different $T$
values is shown in Fig. \ref{diff_prob}. It is seen from Fig.
\ref{diff_prob} that the diffusion probability decreases  very
rapidly (depending on $T$ ) with increasing barrier
 height $H$. At a given $H$, $P_\mathrm{Diffus}$ is larger for higher temperature.

The macroscopic deformation energies are calculated as a function
of the parameter $s$ using the improved version of algebraic
equations \cite{cap2011}. In order to estimate the barrier height,
$H$, $s_{inj}$ is a crucial parameter. In the FBD model this
parameter $s_{inj}$ is a free parameter which is adjusted to
reproduce the measured fusion cross section. In the work by Cap
et al. \cite{cap2011}, $s_{inj}$ has been deduced for 27 cold fusion
reactions including GSI, LBNL and RIKEN data. In that work, the
$s_{inj}$ values are plotted as a function of the  excess of
kinetic energy above the Coulomb barrier, $E_{c.m.} - B_{0}$,
where $B_{0}$ is the mean value of the Coulomb barrier
($V_\mathrm{Coul}$). The overall trend of $s_{inj}$ is of
decreasing nature with increasing $E_{c.m.} - B_{0}$.  It is seen
from Ref. \cite{cap2011} that except the GSI data, all other data
are scattered. For the purpose of present reactions, $s_{inj}$
values for GSI data (from Ref. \cite{cap2011}) are considered and
a linear least-square fit is obtained as shown in Fig.
\ref{sinj_Ecm_B0}, given by:
\begin{eqnarray}
s_\mathrm{inj} = 1.5985 - 0.23587 (E_{c.m.} - B_{0}) ~\mathrm{fm/MeV}.
\label{eq:six}
\end{eqnarray}

Since the present projectile-target nuclei are deformed ones, the
fusion barrier distribution is expected to be quite broad
\cite{wong1973}. Even at $E_{X}< 8$ MeV, a large fraction of the
barrier distribution will have $(E_{c.m.} - B_{0})>30$ MeV, which
will lead to $s_{inj}\sim $ -5 fm as reflected from Fig.
\ref{sinj_Ecm_B0}. For the present reactions,   the barrier
height, $H$ is calculated at $s_{inj}$ = -5, -4, and -3 fm as
shown in  Fig. \ref{H_ZPZT} using the algebraic equations of
macroscopic energies from Ref. \cite{cap2011}. It is seen from
Fig. \ref{H_ZPZT} that the value of $H$   increases with $Z_{P}Z_{T}$ and
it is lower for smaller value of $s_{inj}$. At $s_{inj}$ = -4
fm, $H$ value is around 5.5$\pm$1.5 MeV for all three reactions
considered in the present work (see Table \ref{tab:table1}). In
the estimation of the diffusion probability using Eq.
\ref{eq:five}, the parameters $H$ and $T$ are crucial. At the
excitation energy $E_{X}<$8 MeV, the temperature $T$ is expected
to be $<$1.0 MeV but definitely $>$0.5 MeV. Fig. \ref{diff_prob}
indicates that at $H$ = 5 $\pm$ 1.5 MeV the diffusion probability
will be in between of 10$^{-6}$ and 10$^{-3}$ for
$0.5~\mathrm{MeV} \leq  T \leq 1.0~\mathrm{ MeV}$. Using Eq.
(\ref{eq:two}) and Fig.~\ref{stick} it appears that for the
present reactions, lower limit of  $\sigma_{stick}\times
P_\mathrm{Diffus}$  is $\sim 10^{-7}$ barn.  Present reactions
using the rare-earth nuclei offers a gain factor of the order of
$\sim 10^{4}$ for $\sigma_{stick}\times P_\mathrm{Diffus}$ over
the reactions of cold fusion, as can be seen from Fig. 2 of Ref.
\cite{swiatecki2005}.
\begin{figure*}
\begin{center}
\centering\includegraphics [trim= 0.0mm 0.0mm 0.0mm  0.0mm, angle=360, clip, height=0.29\textheight]{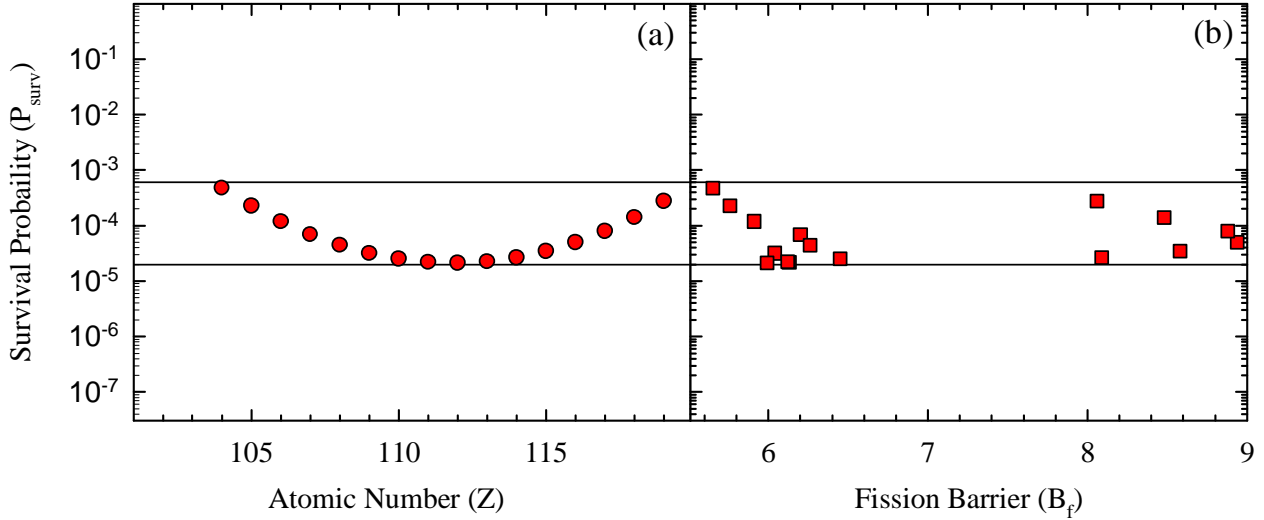}
\caption {(Color online) Survival probability ($P_{surv}$) as derived from Fig. 2 of Ref. \cite{swiatecki2005} as a function of (a) atomic number and 
(b) corresponding fission barrier   of the compound nucleus (see text).  The horizontal bars in each panel  are drawn to indicate the range of $P_{surv}$ from 0.2$\times$10$^{-4}$ to 6$\times$10$^{-4}$. } 
\label{surviv}
\end{center}
\end{figure*}

As far as survival probability ($P_{surv}$) is concerned, in cold fusion reactions when only one neutron is emitted from the compound nucleus, the $P_{surv}$ is the product of probability to emit a neutron rather than fission in the first stage of de-excitation process times the probability $P_{<}$ that the excitation energy (after the emission of neutron) is less than the threshold for second chance fission or second neutron emission:
\begin{eqnarray}
 P_{surv} = \frac{\Gamma_{n}}{\Gamma_{n} +\Gamma_{f}}P_{<}
\label{surv}
\end{eqnarray}

where $\Gamma_{n}$ and $\Gamma_{f}$ are the partial decay widths for first chance neutron emission and  fission, respectively. $P_{surv}$  is expected to be influenced by the properties of the CN such as its ground state mass, its fission barrier, excitation energy, neutron separation energy, shell effects and the level density \cite{swiatecki2005}. As mentioned earlier, a set of cold fusion reactions with $^{208}$Pb and $^{209}$Bi targets has been analyzed with FBD model by Swiatecki et al. \cite{swiatecki2005}. From their work (Fig. 2 of Ref. \cite{swiatecki2005}), we have derived the value of $P_{surv}$ for the elements with atomic numbers $Z$=104 to 119 which are seen to lie in a narrow range of 0.2$\times$10$^{-4}$ to 6$\times$10$^{-4}$. In Fig. \ref{surviv}(a), we show the values of $P_{surv}$ as a function of $Z$ of the compound nucleus. The fission barriers for these compound nuclei vary from 5.5 to 9 MeV \cite{moller_nix_2009}. In Fig. \ref{surviv}(b), we have plotted the value of $P_{surv}$ against the fission barrier taken from Ref. \cite{moller_nix_2009}. It is seen from Figs. \ref{surviv}(a) and (b) that the survival probability has a very weak dependence on the properties of compound nucleus in case of cold fusion reactions (for $Z$=104-119).

The values of $P_{surv}$ can differ in other mass regions, in particular if regions with shell closures are entered like is the
case in the present work. For the present reactions it will be larger than cold as well as hot
fusion reactions due to following reasons: (i) initial excitation
energies can be tuned to be very small ($\sim$8 MeV). Therefore,
shell effects are expected to be more prominent and (ii) good
$n/p$ ratio required for the stability of the super-heavy
elements.

The fission barriers for $Z\geq 120$ have been calculated by different authors of Refs. \cite{moller_nix_2009, koura, sheikh} and are seen to be varying widely between 2 to 8 MeV depending on the model parameters used in their work. However, since the value of $P_{surv}$ for the cold fusion reactions (for $Z$=104-119) corresponding to the excitation energy range of 10 to 15 MeV (below 2n threshold) is not sensitive to the value of fission barrier, we consider a lower limit of 10$^{-4}$ for the survival probability for all the systems considered in the present work. Using this value of $P_{surv}$ and $P_\mathrm{Diffus}$ to be 10$^{-6}$, the lower limit of  final cross sections for the
synthesis of super-heavy nuclei for the present systems having $Z\geq 120$ is arrived to
be  in the range of 1.7$\times 10^{-11}$ barn to 3.0 $\times 10^{-11}$ (see Fig. \ref{stick}) .  Even if we allow  some uncertainties in
the parameter values, the results seem to be quite encouraging.
Present work suggests it to be definitely worth for experimental
investigations using rare-earth nuclear collisions. It is also
necessary to carry out full microscopic calculations to understand
the fusion mechanism for these heavy systems.

\section{\label{sec:level5}Summary}
In the present work, we have made a case for the use of rare-earth
projectile and target nuclei to produce  super-heavy  nuclei in
the range of $Z\sim$120 and above using cold fusion reactions. The advantages offered by these
near symmetric collisions have been outlined. The cross sections
for production of the super-heavy nuclei in these collisions have
been estimated within the framework of the Fusion by Diffusion
model with empirically derived parameter values and are seen to be quite encouraging. It is, however, necessary to carry out
experiments to explore these possibilities of using rare-earth
nuclei in cold fusion reactions for production of super-heavy elements.

\acknowledgments
Authors are thankful to Drs. S. S. Kapoor, A. K. Jain (IITR), and
V. M. Datar for many useful discussions.



\end{document}